\documentclass[preprint,showpacs,preprintnumbers,amsmath,amssymb]{revtex4}
\usepackage{amsmath}

\usepackage{graphicx}
\usepackage{pdfpages}
\usepackage{dcolumn}
\usepackage{bm}
\usepackage{amsmath} 
\usepackage{amsfonts}
\usepackage{amssymb}
\usepackage{url}
\usepackage{microtype}
\usepackage{graphicx}%

\newcommand{\qed}{\nobreak \ifvmode \relax \else
      \ifdim\lastskip<1.5em \hskip-\lastskip
      \hskip1.5em plus0em minus0.5em \fi \nobreak
      \vrule height0.75em width0.5em depth0.25em\fi}

\begin{document}

\preprint{}
\title{Hilbert-Schmidt Separability Probabilities from Bures Ensembles and {\it vice versa}: Applications to Quantum Steering Ellipsoids and Monotone Metrics}
\author{Paul B. Slater}
 \email{paulslater@ucsb.edu}
\affiliation{%
Kavli Institute for Theoretical Physics, University of California, Santa Barbara, CA 93106-4030\\
}
\date{\today}
            
\begin{abstract}
We reexamine a recent analysis in which, using the volume of the associated quantum steering ellipsoid (QES) as a measure, we sought to estimate the probability that a two-qubit state is separable. In the estimation process, we, in effect, sought to attach to states random with respect to Hilbert-Schmidt (HS) measure, the corresponding QES volumes. However, a study of the relations between HS and Bures ensembles and their well-supported separability probabilities of $\frac{8}{33}$ and $\frac{25}{341}$, respectively, now lead us to explore as a possible alternative measure,  the QES volume divided by the $\Pi_{j<k}^{1...4} (\lambda_j-\lambda_k)^2 $ term of the HS volume element (the $\lambda$'s being the four eigenvalues of the associated $4 \times 4$ density matrix $\rho$). This measure is applied to the members of a HS ensemble of random two-qubit states, yielding a QES separability probability estimate of 0.105458. Alternatively, weighting members of a  Bures ensemble by the QES volume divided by the eigenvalue part $\frac{1}{\sqrt{\mbox{det} \rho}}   \Pi_{j<k}^{1...4} \frac{(\lambda_j-\lambda_k)^2}{\lambda_j+\lambda_k}$  of the Bures volume element, gives a close estimate of 0.100223. We also weight members of a HS ensemble by the QES volume divided not only by the indicated HS eigenvalue term, but also by the unitary component $|\Pi_{j<k}^{1...4} \mbox{Re} (U^{-1})  \mbox{Im} (U^{-1})|$ of the volume element. For one hundred thirty (rather variable) independent separability probability estimates, we, then,  obtain median and mean estimates of 0.0447729 and 0.117485 with variance 0.0381468.
\end{abstract}


\maketitle
In a recent study \cite{slater2020quantum}, we reported an estimate of 0.0286 for the probability that a two-qubit state is separable, if one applies to each state, as measure, the volume $V_A$ of the corresponding quantum steering ellipsoid (QES) of (say) Alice.
We have that \cite[eq. (5)]{jevtic2014quantum}
\begin{equation}
\label{VolumeInRho}
V_A = \frac{64\pi }{3}\frac{\left\vert \det  \rho
-\det  \rho^{T_B}  \right\vert }{\left( 1-b^{2}\right) ^{2}}.
\end{equation}
The two-qubit $4 \times 4$ density matrix is denoted by $\rho$  and its partial transpose with respect to Bob's qubit by $\rho^{T_B}$ (obtained by transposing in place the four $2 \times 2$ blocks of $\rho$ \cite[eq. (16.52)]{bengtsson2017geometry}).
$b$ is the norm of Bob's Bloch vector. The  volume $V_B$ of $\mathcal{E}_B$ (the set of states to which Alice can steer Bob, forming an ellipsoid in Bob's Bloch sphere) can be computed from $V_A$ via the  relation
$V_B = \frac{(1-b^2)^2}{(1-a^2)^2}V_A$, where $a$ is the norm of Alice's Bloch vector.

Though not yet presented as a formalized proof, multifaceted numerical and analytical evidence strongly indicates that the probability with respect to Hilbert-Schmidt (HS) measure \cite{zyczkowski2003hilbert} that a member of the fifteen-dimensional convex set of two-qubit systems is separable is $\frac{8}{33} = \frac{2^3}{3 \cdot 11} \approx 0.242424$ \cite{slater2017master,fei2016numerical} \cite[p. 468]{bengtsson2017geometry}. (In fact, Lovas and Andai have given a formal proof that the HS separability probability for two-re[al]bit systems is $\frac{29}{64}$ \cite[Thm. 2]{lovas2017invariance}. Though clearly not so compelling, a highly extensive numerical analysis points to the corresponding separability probability based on the Bures (minimal monotone \cite{sommers2003bures}) measure  being $\frac{25}{341}=\frac{5^2}{11 \cdot 31} \approx 0.0733138$ \cite[sec. II.C.1]{slater2019numerical}. (The HS  qubit-{\it qutrit} separability probability has been conjectured to equal
$\frac{27}{1000} =\frac{3^3}{2^3 \cdot 5^3}= 0.027$ \cite[sec. III.A]{slater2019numerical} \cite[Tab. 1]{khvedelidze2018generation}. \cite[eq.(33)]{milz2014volumes}. A two-qubit separability probability of $1-\frac{256}{27 ]pi^2}$ based on the operator monotone function $\sqrt{x}$ has been reported in \cite[eq. (87)]{slater2017master}--while Lovas and Andai have given an integral formula for the two-rebit case \cite[Thm. 4]{lovas2017invariance}. Any ball with respect to the Bures distance
of a fixed radius in the space of quantum states has the same measure \cite{al2010random}.)

Let us now see if we can  estimate these  conjectured HS and Bures two-qubit separability probabilities-- $\frac{8}{33}$ and $\frac{25}{341}$--in a rather unusual, somehat indirect fashion. The lessons we apparently learn from these exercises, will lead us to reappraise the findings in \cite{slater2020quantum}, suggesting possible alternatives to the estimate of 0.0286 for the QES-based separability probability.

To begin we note that procedures for generating ensembles of quantum states random with respect to the HS and Bures measures have been given in \cite{al2010random}. 
To generate a
HS random $N \times N$ density matrix, one takes a square complex random matrix $A$ of size $N$ pertaining to the
Ginibre ensemble \cite{Gi,Me91} (with real and imaginary parts of
each element being independent normal random variables), and forms 
\begin{equation}
\rho_{\rm HS} \ = \  \frac{\ \ AA^{\dagger}} {{\rm Tr}AA^\dagger} .
\label{hsrand}
\end{equation}
This is by construction, Hermitian, positive definite
and normalised, so it forms a legitimate density matrix.

To generate random Bures states, one takes a complex random matrix $A$ of size $N$ pertaining to the
 Ginibre ensemble and a random unitary matrix $U$
  distributed according to the Haar measure on $U(N)$ \cite{PZK98,Me07}.
Then, one writes down the random matrix
\begin{equation}
\rho_{\rm B} \ = \  \frac{\ \ ({\mathrm 1}+U)AA^{\dagger} ({\mathrm 1}+U^{\dagger})}
{{\rm Tr}[({\mathrm 1}+U)AA^{\dagger} ({\mathrm 1}+U^{\dagger})]} \,.
\label{burrand}
\end{equation}
This is proved to represent a normalized quantum state
distributed according to the Bures measure.

Presumably--based on our indicated previous extensive work \cite{slater2017master,fei2016numerical} \cite[p. 468]{bengtsson2017geometry} \cite[sec. II.C.1]{slater2019numerical}--if we generate a large number of two-qubit states random with respect to Hilbert-Schmidt measure, the sample separability probability will be close to $\frac{8}{33}$, and close to $\frac{25}{341}$ for a Bures ensemble. But, it now appears we  can also obtain such estimated separability probabilities, somewhat indirectly,  using the alternative ensemble.

In fact, employing a Bures ensemble of two-and-a-half million two-qubit density matrices, we have been
able to obtain an estimate of the Hilbert-Schmidt separability probability only 1.00127 as large as $\frac{8}{33}$.
Conversely, using a HS ensemble of four million two-qubit density matrices, we obtained an estimate of the Bures separability probability only 0.990168 as large as $\frac{25}{341}$.

The key to these new findings lies with the use of the Hilbert-Schmidt volume element formula \cite[eq. (15.34)]{bengtsson2017geometry},
\begin{equation} \label{HSvolumeelement}
d V_{\mbox{HS}}=\frac{\sqrt{N}}{2^{(N-1)/2}}   \Pi_{j=1}^{N-1} \mbox{d} \lambda_j \Pi_{j<k}^{1...N} (\lambda_j-\lambda_k)^2  |\Pi_{j<k}^{1...N} \mbox{Re} (U^{-1} \mbox{d}U)_{jk} \mbox{Im} (U^{-1} \mbox{d}U)_{jk} |
\end{equation}
and its Bures counterpart \cite[eq. (15.47)]{bengtsson2017geometry},
\begin{equation} \label{Buresvolumeelement}
d V_{\mbox{Bures}}=\frac{1}{2^{N-1}} \frac{1}{\sqrt{\mbox{det} \rho}}  \Pi_{j=1}^{N-1} \mbox{d} \lambda_j \Pi_{j<k}^{1...N} \frac{(\lambda_j-\lambda_k)^2}{\lambda_j+\lambda_k}  |\Pi_{j<k}^{1...N} \mbox{Re} (U^{-1} \mbox{d}U)_{jk} \mbox{Im} (U^{-1} \mbox{d}U)_{jk} | .
\end{equation}
(Part of the considerable challenge facing us here  is that the QES volume $V_A$ appears to have no ready formulation in terms of the indicated eigenvalue $\lambda$ and unitary $U$ parameters.) 

Now, to estimate the Hilbert-Schmidt separability probability on the basis of a random Bures ensemble--taking ratios of the two volume elements--we attach to each member of the Bures ensemble, the weight
\begin{equation} \label{BuresWeight}
 \Pi_{j<k}^{1...N} (\lambda_j+\lambda_k) \sqrt{\mbox{det} \rho},
\end{equation}
where $\mbox{det} \rho=\Pi_i^{1...N} \lambda_i$. (Since separability probabilities are ratios, scaling factors that are constant across all states--whether entangled or separable--cancel out.)

Conversely, to estimate the Bures separability probability on the basis of a random Hilbert-Schmidt ensemble, we attach to each member of the ensemble, the reciprocal of the weight (\ref{BuresWeight}) (cf. \cite[eq. (11)]{slater2020quantum})
\begin{equation}
\frac{1}{\Pi_{j<k}^{1...N} (\lambda_j+\lambda_k) \sqrt{\mbox{det} \rho}}.
\end{equation}

The indicated numerics certainly indicate that these are appropriate procedures, though a formal argument to this effect still seems to remain open.

Building upon these strong numeric results, leads us now to estimate the QES separability probability by attaching to each member of a random Hilbert-Schmidt ensemble, not simply the QES volume  $V_A$ as in \cite{slater2020quantum}, but the adjusted weight ($N=4$)
\begin{equation} \label{QESHSweight}
  \frac{V_A}{\Pi_{j<k}^{1...N} (\lambda_j-\lambda_k)^2}.
\end{equation}

Employing a Hilbert-Schmidt random ensemble of size ten million, we now obtained a QES separability probability estimate of 0.105458.
(The accompanying Hilbert-Schmidt separability probability  estimate was 0.996472 as large as $\frac{8}{33}$.)

The probability--ensuring entanglement--that $V_A >\frac{4 \pi}{81}$ was estimated as 0.804244, while 0.0902987 of the probability  0.195756 that $V_A <\frac{4 \pi}{81}$ was for entangled states.

Continuing along such lines, now let us seek to  estimate the QES separability probability by attaching to each member of a random Bures ensemble, not the weight $V_A$ as in \cite{slater2020quantum}, but the adjusted weight ($N=4$)
\begin{equation} \label{QESBuresweight}
  \frac{V_A}{\frac{1}{\sqrt{\mbox{det} \rho}}  \Pi_{j<k}^{1...N} \frac{(\lambda_j-\lambda_k)^2}{\lambda_j+\lambda_k}}.
\end{equation}

Employing a Bures random ensemble also of size ten million, we now obtained a QES separability probability estimate of 0.10223 (cf. 0.105458).
The probability--ensuring entanglement--that $V_A >\frac{4 \pi}{81}$ was estimated as 0.812944 
(cf. 0.804244), while 0.0868333 (cf.  0.0902987) of the probability 0.187056 that $V_A <\frac{4 \pi}{81}$  was for entangled states.

\begin{figure}
    \centering
    \includegraphics{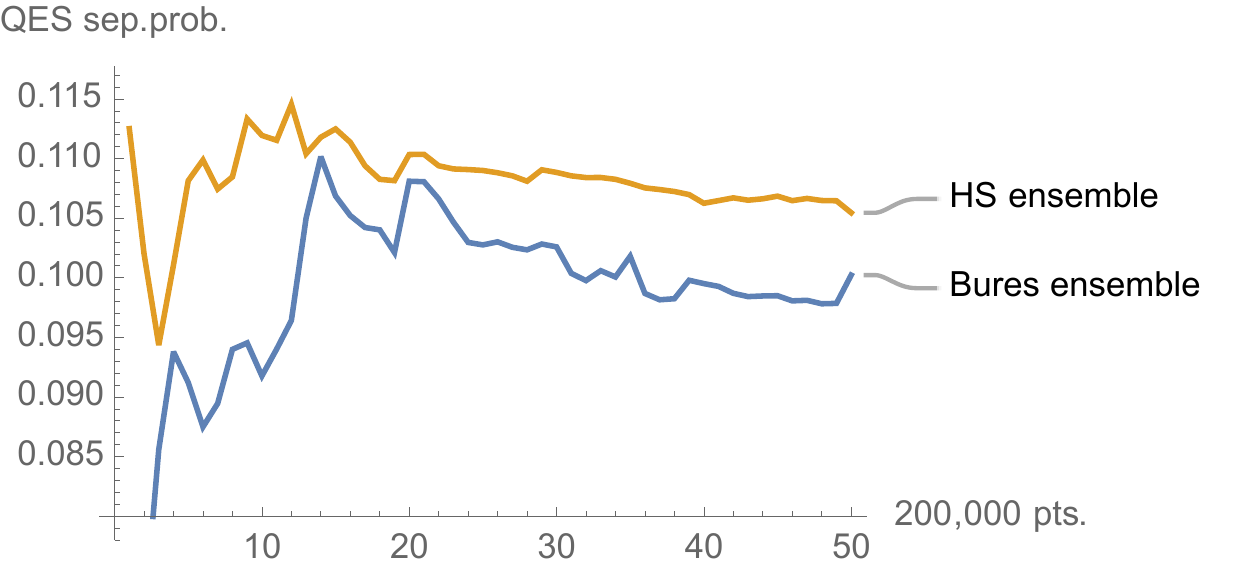}
    \caption{Quantum steering ellipsoid-based estimated separability probability using Hilbert-Schmidt and Bures random ensembles of two-qubit density matrices with weights (\ref{QESHSweight}) and (\ref{QESBuresweight})}
    \label{fig:QESadjustedPlot}
\end{figure}

Figure~\ref{fig:QESadjustedPlot} shows the Bures and Hilbert-Schmidt separability probability estimation process.

Of course, this strategy can immediately be applied with measures other than the QES-based one pursued here--in particular, the vast class of measures based on monotone metrics \cite{petz1996geometries}--including the Kubo-Mori (Bogoliubov), Wigner-Yanase-Dyson, Grosse-Krattenthaler-Slater \cite{grosse2001asymptotic}, geometric and arithmetic average...\cite{petz1996monotone}.
Since each step of the ensemble-generating process requires twice as many normal random variables in the Bures case (since the unitary matrix $U$ is needed), it certainly appears to be more efficient to rely upon Hilbert-Schmidt ensembles. Possible computational reliance upon {\it quasirandom}, as opposed to independently random normal variables, in the ensemble-generating process, was investigated in \cite{slater2020quasirandom}.

\begin{figure}
    \centering
    \includegraphics{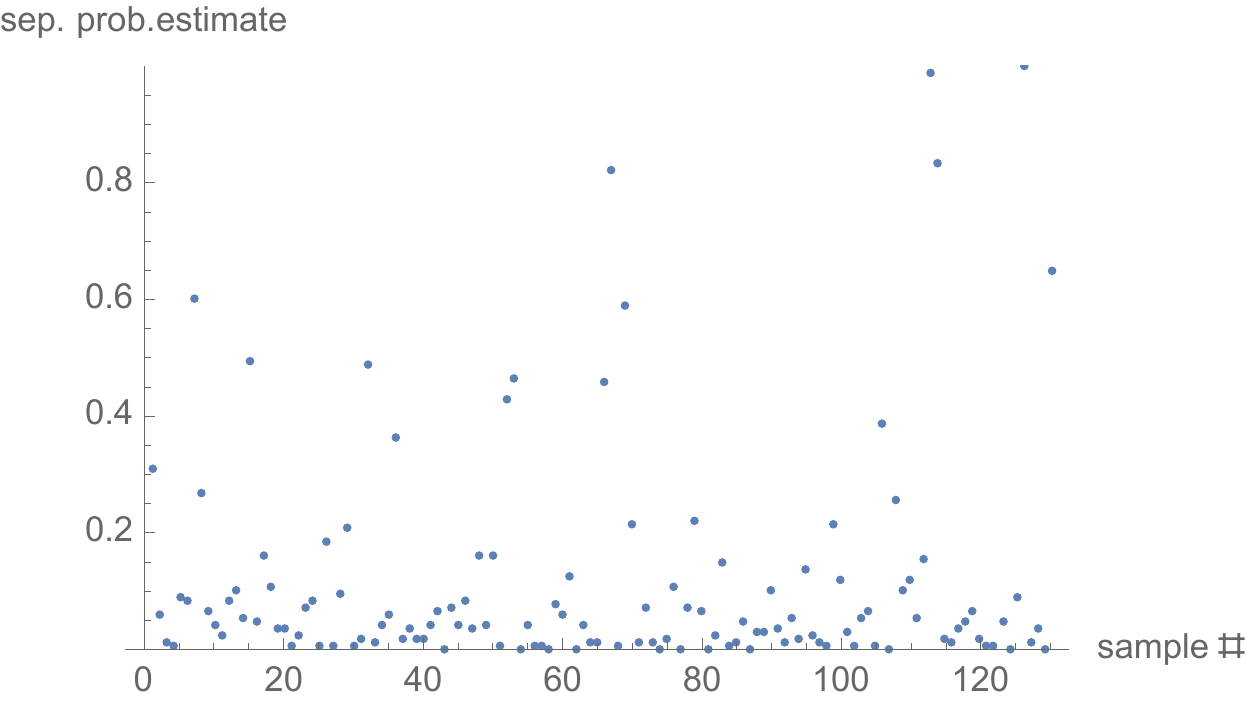}
    \caption{Quantum steering ellipsoid separability probability estimates based on the application of the weight (\ref{revisedHSweight})--now incorporating unitary as well as eigenvalue parameters--to a Hilbert-Schmidt ensemble of size twenty-six million. Each point is an estimate based on two hundred thousand members of the ensemble. The median estimate over the one hundred thirty sets is 0.0447729, while the mean estimate is 0.117485, with variance 0.0381468. The range of estimates is from 0.0000794003 to 0.999997.}
    \label{fig:Unitary}
\end{figure}

It is not fully clear at this point whether or not the most appropriate form of volume element to employ in the case of quantum steering ellipsoids would be of the
product form exhibited by the Hilbert-Schmidt and Bures volume elements (cf. (\ref{HSvolumeelement}, (\ref{Buresvolumeelement})). Given this lack of clarity, we further explored the use as a weight--now incorporating unitary parameters--for members of a Hilbert-Schmidt ensemble (cf.  (\ref{QESHSweight})) of the extended term,
\begin{equation} \label{revisedHSweight}
 \frac{V_A}{   \Pi_{j<k}^{1...4} (\lambda_j-\lambda_k)^2  |\Pi_{j<k}^{1...4} \mbox{Re} (U^{-1})  \mbox{Im} (U^{-1}) |}  .
\end{equation}
Then, employing twenty-six million members of such an ensemble, partitioned into groups of two hundred thousand, we obtained rather widely varying estimates  for the QES-based separability probability (Fig.~\ref{fig:Unitary}). 

Attaching to seven million members of a Hilbert-Schmidt ensemble, the corresponding volume $V_A$ of Alice's quantum steering ellipsoid, the average value of $V_A$ was estimated to 
be 0.20703 that ($\frac{4 \pi}{3}$) of the Bloch (unit) sphere within which the ellipsoids must lie. 
\begin{acknowledgements}
This research was supported by the National Science Foundation under Grant No. NSF PHY-1748958.
\end{acknowledgements}

\bibliography{main}

\end{document}